\documentclass[aps,prc,twocolumn,showpacs]{revtex4}
\usepackage{amssymb}

\usepackage{amsmath,amssymb,mathrsfs}
\usepackage{graphicx}
\usepackage{amsthm}
\usepackage{amscd}
\usepackage{bm}
\usepackage{dsfont}

\usepackage{lscape}

\begin{document}

\title{The probability distribution of the number of electron-positron \\
pairs produced in a uniform electric field}
\author{M. I. Krivoruchenko}
\affiliation{Institute for Theoretical and Experimental Physics \\
B. Cheremushkinskaya 25$\mathrm{,}$ 117218 Moscow$\mathrm{,}$ Russia and \\
Department of Nano-$\mathrm{,}$ Bio-$\mathrm{,}$ Information and Cognitive Technologies\\
Moscow Institute of Physics and Technology \\
9 Institutskii per.$\mathrm{,}$ 141700 Dolgoprudny$\mathrm{,}$ Russia
}

\begin{abstract}
The probability-generating function of the number of electron-positron pairs produced 
in a uniform electric field is constructed.
The mean and variance of the numbers of pairs are calculated, and analytical
expressions for the probability of low numbers of electron-positron pairs are given. 
A recursive formula is derived for evaluating the probability of any number of pairs. 
In electric fields of supercritical strength $|eE| > \pi m^2/ \ln 2$, 
where $e$ is the electron charge, $E$ is the electric field, and $m$ is the electron mass, 
a branch-point singularity of 
the probability-generating function penetrates the unit circle $|z| = 1$,
which leads to the asymptotic divergence of the cumulative probability.
This divergence indicates a failure of the continuum limit approximation.
In the continuum limit and for any field strength, the positive definiteness of the probability 
is violated in the tail of the distribution.
Analyticity, convergence, and positive definiteness are restored upon the summation over 
discrete levels of electrons in the normalization volume.
Numerical examples illustrating the field strength dependence of the asymptotic behavior of the probability distribution 
are presented.
\end{abstract}


\maketitle
%%%%%%%%%%%%%%%%%%%%%%%%%%%%%%%%%%%%%%%%%%%%%%%%%%%%%%%%
\section{Introduction}
%%%%%%%%%%%%%%%%%%%%%%%%%%%%%%%%%%%%%%%%%%%%%%%%%%%%%%%%

Klein's paradox \cite{KLEI29} is an excellent illustration of the problems
associated with the existence of solutions with negative energy in the Dirac
equation. Since the formulation of the paradox, it has attracted considerable attention. 
The physical meaning of the apparently anomalous behavior
of the solutions of the Dirac equation has gradually been clarified,  
and the complete solution to the paradox has finally been obtained \cite{SAUT31,HEIS36,HUND40,SCHW51,NIKI69}. 
The effect is related to the production of
electron-positron pairs in an external electric field or, equivalently, with the possibility of the tunneling
of electrons from the region containing solutions with negative energy to the region
containing solutions with positive energy. 

Verification of the Schwinger mechanism of electron-positron pair production \cite{SCHW51,NIKI69,NIKI70,NARO70,LEBE84,DITT85} 
is still an open experimental challenge, due to the smallness of the effect 
for the electric fields that are available in a modern laboratory.
In recent years, the mechanism of pair production based on multiple soft-photon absorption has been discussed.
The combination of a strong, slowly varying field and a weak but dynamical field lowers the threshold for pair production, 
which provides a possible method to create electron-positron pairs 
in experiments with ultra-high-intensity lasers \cite{SCHU08}.
Various other possibilities to create electron-positron pairs in laser beams 
are also discussed \cite{RING01,NARO04,DUNN09,HEBE09}. 
Latest developments in this field are reviewed in Ref.~\cite{DUNN10}. 

The probabilities of creating one, two, or more electron-positron pairs 
that contain detailed information about the pair-production mechanism 
fundamentally allow for a direct experimental measurement.
In this paper, we construct the probability distribution of the electron-positron pairs produced
in a uniform static electric field. In Sec. II, the probability distribution 
is calculated %in the continuum limit approximation 
on the basis of the probability-generating function (PGF), and in Sec. III, 
the probability distribution is calculated with the help of recursion relations. 
As shown in Sec. IV, the continuum limit approximation is restricted 
by weak electric fields. In strong electric fields, the analytic properties of the PGF are violated 
and the cumulative probability is divergent. We show that the summation over discrete modes restores both
the analyticity of the PGF and the convergence of the series. 
The continuum limit approximation also leads %, for any field strength, 
to a breakdown of the positive definiteness 
of the probability in the tail of the distribution. 
This shortcoming is also overcome in a discrete scheme. 
The main features of the asymptotic behavior 
of the probability distribution are illustrated with numerical examples. Final conclusions are collected in Sec. V.

%%%%%%%%%%%%%%%%%%%%%%%%%%%%%%%%%%%%%%%%%%%%%%%%%%%%%%%%
\section{The probability-generating function}
%%%%%%%%%%%%%%%%%%%%%%%%%%%%%%%%%%%%%%%%%%%%%%%%%%%%%%%%

In the standard formulation of the problem, 
the electric field is switched on adiabatically for a time $T$, and then the field is adiabatically switched off. 
The probability that a vacuum remains a vacuum \cite{SCHW51} and the average number of produced pairs \cite{NIKI69} 
are of interest. We are also interested in the probability distribution of the number of produced pairs.

Electrons with negative energy moving against the electric field lines 
are slowed down and stopped, and they either tunnel with a probability of $w_{\alpha }$ into positive-energy states 
or turn back with a probability of $1-w_{\alpha }$, 
in which case they are further accelerated along the field lines. 
We denote by $\alpha =(\mathbf{p},\sigma )$ the quantum numbers of electrons, 
where $\mathbf{p}$ is the initial momentum, and $\sigma $ is the electron spin projection. 
In a uniform electric field, the semiclassical expression
for the probability of tunneling appears to be accurate:
\begin{equation}
w_{\alpha }=\exp \left(-\frac{\pi (m^{2}+\mathbf{p}_{\bot }^{2})}{|eE|}\right),
\label{PROB}
\end{equation}
where $e$ is the electron charge, $E$ is the electric field strength, and
$\mathbf{p}_{\bot }$ is the electron momentum orthogonal to the field.

Thus, we have an example of Bernoulli trials:
the tunneling of electrons occurs with a probability of $w_{\alpha }$ for each electron, 
and the reflection of electrons occurs with a probably of $1 - w_{\alpha }$ for each electron.

The vacuum-to-vacuum transition probability equals 
\begin{equation*}
p_{0}=\prod_{\alpha }(1-w_{\alpha }). 
\end{equation*}
In electric fields, any number of electron-positron pairs can be produced. 
The sum of all of the probabilities is equal to unity:
\begin{eqnarray}
&&\prod_{\alpha }(1-w_{\alpha })+\sum_{\alpha }w_{\alpha }\prod_{\beta \neq
\alpha }(1-w_{\beta }) \nonumber \\
&+&\frac{1}{2!}\sum_{\alpha \neq \beta }w_{\alpha}w_{\beta } \prod _{\gamma \neq \alpha , \beta}
(1-w_{\gamma })+\ldots =1.  \label{UNIT}
\end{eqnarray}
Whenever the series converges, this algebraic identity is independent of $w_{\alpha}$.
The proof of this fact can easily be made using a representation of the series in the form of a product, namely,
\begin{equation}
p_{0}\prod_{\alpha }\left( 1+\nu_{\alpha } \right), \label{prod}
\end{equation}
where $\nu_{\alpha } = w_{\alpha}/(1-w_{\alpha})$. 
The sequential terms of the left-hand side of Eq.~(\ref{UNIT}) have the meaning of probability 
that there are $n=0,1,2,...$ electron-positron pairs.

Consider a more general expression:
\begin{eqnarray}
\varphi (z)&=& p_{0}\prod_{\alpha }\left( 1+z\nu_{\alpha }\right).  \label{CHAR}
\end{eqnarray}
In probability theory and statistics, the function $\varphi (z)$ 
is known as the PGF. 
A stochastic process to which it corresponds is characterized by the probability distribution
\begin{equation}
p_{n}=\frac{1}{n!}\frac{d^{n}}{dz^{n}}\varphi (z)|_{z=0}.  \label{PRON}
\end{equation}
In our case, each of the quantities $p_{n}$ describes the probability 
of there being $n$ electron-positron pairs.

A PGF has many useful properties.
In particular, $\varphi (z) = \mathrm{E}[z^n]$ is the expectation value of $z^{n}$; 
the equation $\varphi (1)=1$ is equivalent to Eq.~(\ref{UNIT}); 
$\varphi (0)=p_{0}$ is the vacuum-to-vacuum transition probability; 
the average number of electron-positron pairs is $\mathrm{E}[n]=\varphi^{\prime}(1)$; 
the variance equals $\mathrm{V}[n]= \mathrm{E}[n^2] - \mathrm{E}[n]^2 =
\varphi^{\prime \prime }(1)+\varphi ^{\prime }(1)-\varphi ^{\prime }(1)^{2}$. 

The logarithm of $\varphi (z)$ includes a summation over the indices $\alpha$.
During the period of observation $T$, the relativistic electrons with longitudinal momenta
$ 0 < p_{\parallel} < | eE | T $ slow down and then stop their movement along the field lines. 
Once they have stopped, these electrons acquire a chance to tunnel. 
In the continuum limit approximation, computing the discrete sum is replaced by integrating over the phase space:
\begin{equation}
\sum_{\alpha } \ldots \rightarrow \int \frac{2Vd^{3}p}{(2\pi )^{3}}\ldots = VT|eE|\int 
\frac{2d^{2}p_{\bot }}{(2\pi )^{3}}\ldots , 
\label{SUBS}
\end{equation}
where $V$ is the normalization volume. 
The probability of tunneling does not depend 
on $p_{\parallel}$, so this component was integrated out.

Using well-known arguments \cite{SCHW51,NIKI69,NIKI70}, we obtain 
\begin{equation}
\varphi (z)=\exp \left( -\gamma \mathrm{Li_2}( (1-z)\xi ) \right) ,  
\label{SCHW}
\end{equation}
where $\mathrm{Li_2}(t)$ is the dilogarithm
\begin{equation*}
\mathrm{Li_2}(t) =-\int\limits_{0}^{t}\frac{\ln (1-\theta )}{\theta }d\theta
\end{equation*}
and
\begin{equation*}
\gamma  = VT\frac{e^{2}E^{2}}{4\pi ^{3}}, \;\;\;\; \xi = \exp (-\frac{\pi m^{2}}{|eE|}).
\end{equation*}
The value of $\gamma$ is the effective number of negative-energy electrons 
that reach a turning point during time $T$ and then obtain a chance to tunnel.
To  replace the discrete sum with an integral, one has to require $\gamma \gg 1$. 
As we shall see later, this condition, although necessary, 
is not sufficient.
\footnote{
For charged bosons, analogous arguments lead to the probability-generating function of the form (\ref{SCHW}) 
with the substitution $\textrm{Li}_2((1-z)\xi) \to -\textrm{Li}_2(-(1-z)\xi) $. 
The factor $\gamma$ must be modified according to the number of degenerate states of bosons with momentum $\mathbf{p}$.
}

The vacuum-to-vacuum transition probability and the probabilities of producing one, two, or three electron-positron pairs 
can easily be found by means of Eq.~(\ref{PRON}): 
%\begin{widetext}
\begin{eqnarray}
p_{0} &=&\exp \left( -\gamma \mathrm{Li_2}(\xi)\right) ,  \label{p1} \\
p_{1} &=&\gamma Lp_{0},  \label{p2} \\
p_{2} &=&\frac{\gamma}{2}\left( \gamma L^{2}+ L - \frac{ \xi }{1-\xi }\right) p_{0},  \label{p3} \\ 
p_{3} &=&\frac{\gamma}{6}\left( \gamma^{2}L^{3}
                                 +3 \gamma    L^{2}
                                 -\frac{-2+2\xi+3\gamma \xi}{1-\xi }L \right. \nonumber \\
                                 &&\;\;\;\;\;\;+\left. \frac{\xi (3\xi -2)}{(1-\xi )^{2}}\right) p_{0},  \label{p4}
\end{eqnarray}
%\end{widetext}
where $L  = -\ln (1-\xi )$. The zero-order term $p_{0}$ is in agreement with the result of Schwinger \cite{SCHW51}.
The average number of pairs, which was first found by Nikishov \cite{NIKI69}, equals
\begin{equation}
\mathrm{E}[n] = \gamma \xi ,  \label{AVER}
\end{equation}
while the variance in the pair number is given by 
\begin{eqnarray}
\mathrm{V}[n]  = \frac{1}{2}\gamma \xi (2-\xi ).  \label{DISP}
\end{eqnarray}

The leading terms of $p_{n}/p_0 \sim (\gamma L)^n /n! $ induce the Poisson distribution. 
They describe the $n$ uncorrelated pairs and, 
in violation of fermion statistics, they count configurations with two or more negative-energy electrons in the same state. 
The lower-order terms in $\gamma$ remove these unphysical contributions. 
For example, the probability of there being two pairs, $p_2$, is given by
\begin{equation}
\frac{1}{2}\sum_{\alpha \neq \beta} \nu_{\alpha }\nu_{\beta } p_0 = 
\frac{1}{2}(\sum_{\alpha } \nu_{\alpha })^2  p_0 - \frac{1}{2} \sum_{\alpha } (\nu_{\alpha })^2 p_0.
\label{exam}
\end{equation}
The first positive term equals $p_0(\gamma L)^2/2!$;
the second negative term has the order $O(\gamma)$ (when it is divided by $p_0$). 
The left-hand side is positive, so the second term is smaller than the first one. 
To avoid violating this condition, the value of $\xi$ should not be especially close to unity:
\begin{equation}
1 - \xi \gtrsim  1/\gamma. 
\label{REST}
\end{equation}

The expansion in the powers of $\gamma$ in the expressions for $p_n/p_0$ 
arises due to fermion statistics, and this expansion has a combinatorial meaning.
%The correlation effects in the pair production are discussed in Ref.~\cite{LEBE84}. 
The production of uncorrelated pairs is described by the Poisson distribution 
with the mean value $\mathrm{E}[n] = \gamma L$.  
The correlation effects show up in the deviation of $\mathrm{Li}_2(\xi)$ from $L$ in the exponent of $p_0$,
in the deviation of $\xi$ from $L$ in the mean value, etc.
The pre-exponential factor of $p_1$ has a purely Poisson form, since in the one-pair production
the correlations are not important.

%%%%%%%%%%%%%%%%%%%%%%%%%%%%%%%%%%%%%%%%%%%%%%%%%%%%%%%%%%%%%%%%%%%%%%%%%%%%
%%%%%%%%%%%%%%%%%%%%%%%%%%%%%%%%%%%%%%%%%%%%%%%%%%%%%%%%%%%%%%%%%%%%%%%%%%%%
%%%%%%%%%%%%%%%%%%%%%%%%%%%%%%%%%%%%%%%%%%%%%%%%%%%%%%%%%%%%%%%%%%%%%%%%%%%%
\vspace{-4 mm}
\begin{center}
\begin{figure}[t] %[!htb] %[h] %
\includegraphics[angle = 0,width=0.450\textwidth]{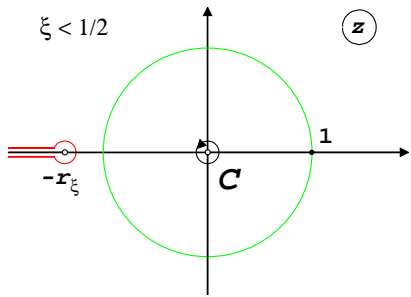}
%\vspace{-5 mm}
\caption{(color online)
The complex $z$-plane of the PGF for electric fields of subcritical strength. 
The branch-point singularity of the PGF at $z= -r_{\xi}$ is outside the unit circle $|z|=1$. 
The contour $C$ shows the path of integration in Eq.~(\ref{COUN}); 
the integrand also has a pole singularity at $z=0$.   
}
\label{fig1a}
\end{figure}
\end{center}
\vspace{-6mm}
%%%%%%%%%%%%%%%%%%%%%%%%%%%%%%%%%%%%%%%%%%%%%%%%%%%%%%%%%%%%%%%%%%%%%%%%%%%%
%%%%%%%%%%%%%%%%%%%%%%%%%%%%%%%%%%%%%%%%%%%%%%%%%%%%%%%%%%%%%%%%%%%%%%%%%%%%
%%%%%%%%%%%%%%%%%%%%%%%%%%%%%%%%%%%%%%%%%%%%%%%%%%%%%%%%%%%%%%%%%%%%%%%%%%%%

%%%%%%%%%%%%%%%%%%%%%%%%%%%%%%%%%%%%%%%%%%%%%%%%%%%%%%%%
\section{Recursion relations}
%%%%%%%%%%%%%%%%%%%%%%%%%%%%%%%%%%%%%%%%%%%%%%%%%%%%%%%%

The calculation of $p_{n}$ for large values of $n$ can be performed recursively using 
the method employed in Ref. \cite{RADU12}.

\begin{center}
\begin{figure}[t] %[!htb] %[h] %
\includegraphics[angle = 0,width=0.450\textwidth]{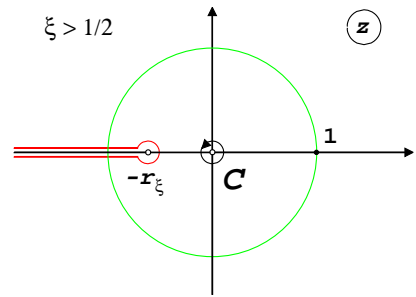}
%\vspace{-5 mm}
\caption{(color online)
The complex $z$-plane of the PGF for electric fields of supercritical strength.
The branch-point singularity of the PGF at $z= -r_{\xi}$ 
penetrates the unit circle $|z|=1$, which destroys the convergence of the cumulative probability.  
The contour $C$ shows the path of integration in Eq.~(\ref{COUN}). 
}
\label{fig1b}
\end{figure}
\end{center}

The function $ \mathrm{Li_2} (t) $ is analytic in the complex $t$-plane, except for the logarithmic branch point at $t= 1$. 
The function $ \varphi (z) $, which is expressed in terms of $ \mathrm{Li_2}(t) $, also
has a logarithmic branch point at $z = - r_{\xi} \equiv 1 - 1/ \xi $. 
For $ \xi <1 $, $ \varphi (z) $ is analytic in the neighborhood of $ z = 0 $; therefore, we can represent $ p_{n} $ 
by the following contour integral:
\begin{equation}
p_{n}=\frac{1}{2\pi i}\int\limits_{C}\frac{dz}{z^{n+1}}\varphi (z).
\label{COUN}
\end{equation}
The contour $C$ encompasses the point $z=0$, as shown in Figs.~\ref{fig1a} and \ref{fig1b}.
As long as $\varphi (z)$ is regular at $z=0$, $p_{n}=0$ for $n=-1,-2,\ldots .$

%%%%%%%%%%%%%%%%%%%%%%%%%%%%%%%%%%%%%%%%%%%%%%%%%%%%%%%%%%%%%%%%%%%%%%%%%%%%
%%%%%%%%%%%%%%%%%%%%%%%%%%%%%%%%%%%%%%%%%%%%%%%%%%%%%%%%%%%%%%%%%%%%%%%%%%%%
%%%%%%%%%%%%%%%%%%%%%%%%%%%%%%%%%%%%%%%%%%%%%%%%%%%%%%%%%%%%%%%%%%%%%%%%%%%%
\vspace{-4 mm}
\vspace{-6mm}
%%%%%%%%%%%%%%%%%%%%%%%%%%%%%%%%%%%%%%%%%%%%%%%%%%%%%%%%%%%%%%%%%%%%%%%%%%%%
%%%%%%%%%%%%%%%%%%%%%%%%%%%%%%%%%%%%%%%%%%%%%%%%%%%%%%%%%%%%%%%%%%%%%%%%%%%%
%%%%%%%%%%%%%%%%%%%%%%%%%%%%%%%%%%%%%%%%%%%%%%%%%%%%%%%%%%%%%%%%%%%%%%%%%%%%

To obtain the recursion, we carry out integration by parts in Eq.~(\ref{COUN}): 
\[
p_{n}=\frac{1}{2\pi i}\int\limits_{C}\frac{1}{nz^{n}}\varphi ^{\prime
}(z)dz. 
\]
This expression also follows directly from the definition of PGF.
The first derivative of $\varphi (z)$ has a simple form:
\begin{equation}
\varphi ^{\prime }(z)=-\frac{\gamma }{1-z}\ln \left( 1-(1-z)\xi \right)
\varphi (z).  \label{WPRI}
\end{equation}
By decomposing the logarithm in front of $\varphi (z)$ first
in a power series of $1-z$, then in a power series of $z$, we obtain 
\[
p_{n}=\frac{\gamma }{n}\sum_{k=1}^{\infty }\frac{\xi ^{k}}{k}%
\sum_{l=0}^{k-1}(-1)^{l}\left( 
\begin{array}{c}
k-1 \\ 
l
\end{array}
\right) p_{n-l-1}. 
\]
One can carry out the summation explicitly: 
\begin{eqnarray}
g_{l} &=& (-1)^{l}\sum_{k=l+1}^{\infty }\frac{\xi ^{k}}{k}\left( 
\begin{array}{c}
k-1 \\ 
l
\end{array}
\right) \nonumber \\
&=&(-1)^{l}\frac{\xi ^{l+1}}{l+1}F(l+1,l+1;l+2;\xi ),
\label{GELL}
\end{eqnarray}
where $F(a,b;c;z)$ is the hypergeometric function. 
For $l = 0$, this expression gives $g_{0} = L$; for $l \geq 1$ we obtain
\begin{equation*}
g_{l} = L + \sum_{m=1}^{l}(-1)^m \frac{1}{m r_{\xi}^m}.
\end{equation*}

Finally, one finds 
\begin{equation}
p_{n}=\frac{\gamma }{n}\sum_{l=0}^{n-1}g_{l}p_{n-l-1}.
\label{RECU}
\end{equation}
This equation gives the recursive definition of the probability distribution 
$p_{n}$ of the random variable, $n$, which represents the number of electron-positron pairs produced in a uniform
electric field. 

Because 
\[
\sum_{l=0}^{\infty }g_{l}=\xi , 
\]
we reproduce Eq.~(\ref{AVER}) after the summation of $np_{n}$ over the pair number.

%%%%%%%%%%%%%%%%%%%%%%%%%%%%%%%%%%%%%%%%%%%%%%%%%%%%%%%%
\section{Asymptotic behavior of the probability distribution}
%%%%%%%%%%%%%%%%%%%%%%%%%%%%%%%%%%%%%%%%%%%%%%%%%%%%%%%%

The radius of convergence of the Taylor series expansion of $ \varphi (z) $ at $z=0$ is determined 
by the distance $r_{\xi}$ to the nearest singularity. The series over $z$ allows one to evaluate 
$ \varphi (z) $ at $z=1$, provided $r_{\xi} = 1/\xi - 1$ is greater than one.
This condition is satisfied for $ \xi < 1/2 $. 
Under this condition, one can rigorously prove that the sum of the probabilities (\ref{PRON}) equals one. 

The appearance of the branch-point singularity inside the unit circle destroys 
the precise probabilistic meaning of the PGF.
The only explanation we can offer for this phenomenon is that the infinite product (\ref{prod}) 
diverges in the continuum limit. 
If this divergence did not occur, the radius of convergence would be at least 1. 
Although product (\ref{prod}) diverges, product (\ref{CHAR}), 
which differs from (\ref{prod}) only by a factor of $z$ for $\nu_{\alpha}$, 
converges for $z$ inside the circle $|z| < r_{\xi} < 1$. 
If $1 < r_{\xi}$, the first product converges; the second product converges 
inside the circle $|z| < r_{\xi}$.
Thus, although the $p_n$ of equation (\ref{PRON}) are analytic functions of the parameter $\xi$, 
and although equation (\ref{PRON}) permits an analytical continuation to the region $\xi > 1/2$, 
the physical meaning of the $p_n$ in strong fields is uncertain.
Evidently, the applicability of equation (\ref{PRON}) is limited by the condition $\xi < 1/2$.

The effective number of electrons with negative energy that have a chance to tunnel 
was previously estimated as $\sim \gamma$. Therefore, it is expected that the $p_n$ are vanishingly 
small for $n \gtrsim  \gamma$. We also have a condition that the term ${(\gamma L)^{n}}/{n!}$ 
in the expansion of $p_n/p_0$ over $\gamma$ is the leading one, and therefore it obviously exceeds the lowest-order term 
$\sim \gamma$. The estimate of the latter gives 
(cf. Eqs.~(\ref{p2}) - (\ref{p4}) and (\ref{exam}))
\begin{equation}
\Delta p_n /p_0 \sim (-1)^{n-1}\sum_{\alpha } (\nu_{\alpha })^n \sim (-1)^{n-1}/r_{\xi}^{n - 1}.
\label{ILLU}
\end{equation}
The condition $|\Delta p_n |/p_0 \lesssim {(\gamma L)^{n}}/{n!}$
is equivalent to 
\begin{equation}
n \lesssim n_c = \gamma L r_{\xi}. 
\label{CONS}
\end{equation}
This inequality is fulfilled for all $n \lesssim \gamma$, so, again, we find $r_{\xi} \gtrsim 1$,
which is in agreement with the previous analysis. For $n = 2$, we return to the less stringent constraint (\ref{REST}).
The condition $r_{\xi} > 1$ also provides a term-by-term convergence of the sum of $p_n$.
For large $n$, the inequality (\ref{CONS}) is violated in any case. This violation does not lead to contradictions 
as long as the sum of the $p_n$ is saturated for $n \lesssim \gamma$. 

Apparently, the reason that the $p_n$ do not sum up to unity for $r_{\xi} < 1$  is 
that the replacement (\ref{SUBS}) is not justified.
The subtraction of terms, which restores fermion statistics, gives rise to contributions 
of the type (\ref{ILLU}).
For $n \sim \gamma$, the crucial  point is whether the magnitude of $\nu_{\alpha }$ 
is greater than unity or less than unity. 
If $\nu_{\alpha }<1$, which corresponds to $w_{\alpha} < 1/2$ and $\xi < 1/2$, the correction is small. 
This is the case depicted in Fig.~\ref{fig1a}.
If $\xi > 1/2$, the long-wavelength modes 
$0 < p_{\bot}^2 < \ln 2 |eE|/\pi - m^2$ give a large contribution that 
can only be evaluated in a discrete scheme. 
This case is shown in Fig.~\ref{fig1b}.
The sum over the discrete modes provides 
a delicate cancellation between large correlation terms.

The infinite product (\ref{prod}) converges if it is calculated over the discrete modes.
In order to prove this assertion, we proceed with the combinatorial representation of $p_n$:
\[
p_{n} = \frac{1}{n!}\sum_{\alpha_{1}\ldots \alpha_{n}}(\prod\limits_{i=1}^{n}\nu_{\alpha_{i}}) p_{0}.
\]
The sum here extends over all ordered sets of the $n$ pairwise-distinct indices $\alpha_{i}$. 
The inclusion in the sum of diagonal terms $\alpha_{i} = \alpha_{j}$ increases the value of $p_n$. 
As an upper bound we have the estimate 
$p_{n} < \bar{p}_{n} = p_{0}(\gamma \tilde{L} )^{n}/n!$, 
where 
\[
\gamma \tilde{L} = \sum_{\alpha }\nu_{\alpha }.
\]
This series converges regardless of whether or not $\xi$ is small.
The cumulative probability appears to be bounded:
\[
\sum_{n=0}^{\infty }p_{n} < \sum_{n=0}^{\infty }\bar{p}_{n}=\exp (\gamma \tilde{L}) p_0.
\]
The infinite product (\ref{prod}) converges; therefore, 
the $p_n$ are summed up in the unity, 
by virtue of Eq.~(\ref{UNIT}), for any field strength.
It follows that $1 < \exp (\gamma \tilde{L}) p_0$. We remark that this inequality is fulfilled 
in the continuum limit where $ \tilde{L} = L > \mathrm{Li}_2(\xi)$.

%%%%%%%%%%%%%%%%%%%%%%%%%%%%%%%%%%%%%%%%%%%%%%%%%%%%%%%%%%%%%%%%%%%%%%%%%%%%
%%%%%%%%%%%%%%%%%%%%%%%%%%%%%%%%%%%%%%%%%%%%%%%%%%%%%%%%%%%%%%%%%%%%%%%%%%%%
%%%%%%%%%%%%%%%%%%%%%%%%%%%%%%%%%%%%%%%%%%%%%%%%%%%%%%%%%%%%%%%%%%%%%%%%%%%%
%\vspace{-4 mm}
\begin{center}
\begin{figure}[t] %[!htb] %[h] %
\includegraphics[angle = 0,width=0.450\textwidth]{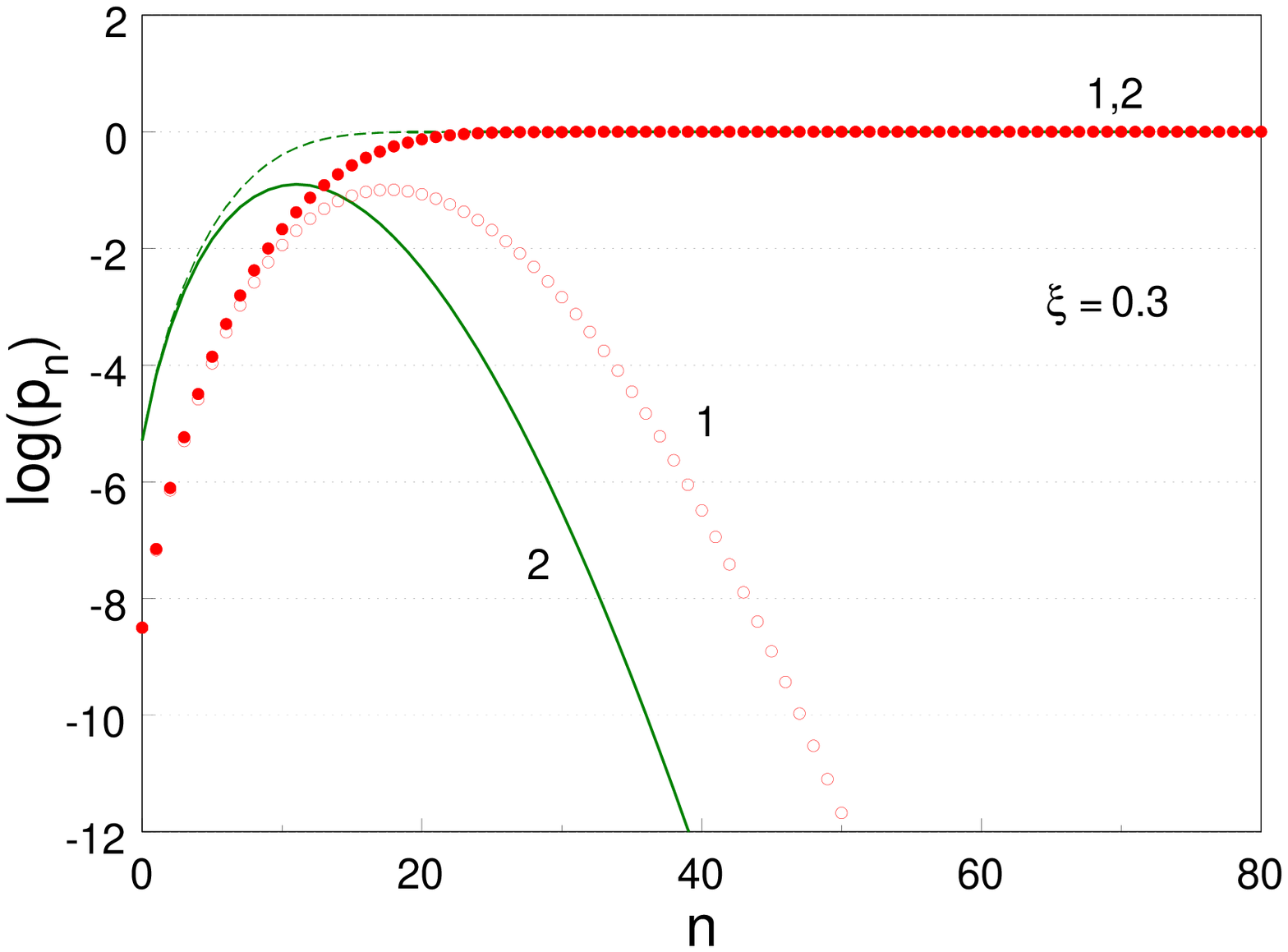}
\caption{(color online)
The probability distribution (the solid curve and empty circles) 
and the cumulative probability distribution (the dashed curve and solid circles) 
of the number of electron-positron pairs, $n$, 
produced in a uniform electric field of subcritical strength $\xi = 0.3$ 
for $\gamma_{\Vert } = 10$, $\gamma _{\bot } = 6$, and $\gamma = 60$. 
Circles 1 show the continuum limit, and curves 2 correspond to the discrete electron levels.    
The series of $p_n$ converge in both cases. The distributions are plotted in the $\log_{10}$ scale.
}
\label{fig1}
\end{figure}
\end{center}
\vspace{-8mm}
%%%%%%%%%%%%%%%%%%%%%%%%%%%%%%%%%%%%%%%%%%%%%%%%%%%%%%%%%%%%%%%%%%%%%%%%%%%%
%%%%%%%%%%%%%%%%%%%%%%%%%%%%%%%%%%%%%%%%%%%%%%%%%%%%%%%%%%%%%%%%%%%%%%%%%%%%
%%%%%%%%%%%%%%%%%%%%%%%%%%%%%%%%%%%%%%%%%%%%%%%%%%%%%%%%%%%%%%%%%%%%%%%%%%%%

It is known that limiting procedures modify the analytic properties of functions.
In our case, an infinite sequence of zeros in the PGF (\ref{CHAR}) has been transmuted in a branch cut.
The non-analyticity is thus associated with 
the replacement of summation over $\alpha$ by the phase space integral.
This remark, incidentally, describes how to restore analyticity: 
We have, unsurprisingly, to refuse performing calculations in a continuous scheme.
This prescription is in agreement with the earlier argument based on the need 
to have the cancellation of large correlation terms, which can be achieved only in a discrete scheme.

One can assume that, in Eq.~(\ref{CONS}), $n_c$ reveals the maximum value of $n$ for which
the calculation of $ p_n $ in the continuum limit is justified for $r_{\xi} < 1$.
If this conjecture is correct, then expression (\ref{p1}) could be valid for arbitrarily small $r_{\xi}$.

As an illustration, we consider a numerical example of the probability distribution 
both in weak and strong electric fields 
in the continuum limit and for a discrete set of levels. The momentum is quantized, 
as in the nonrelativistic case for an infinite potential wall with $p = \pi n/L$, where
$n$ is a positive integer, $L$ is the size of normalization box, and the volume equals $V=L^3$. 
The boundary conditions for the Dirac equation have a more complicated form. 
Our goal here is to clarify the principal distinction between the continuum limit and a discrete 
scheme; therefore, the relativistic modification of the boundary conditions is neglected.

%%%%%%%%%%%%%%%%%%%%%%%%%%%%%%%%%%%%%%%%%%%%%%%%%%%%%%%%%%%%%%%%%%%%%%%%%%%%
%%%%%%%%%%%%%%%%%%%%%%%%%%%%%%%%%%%%%%%%%%%%%%%%%%%%%%%%%%%%%%%%%%%%%%%%%%%%
%%%%%%%%%%%%%%%%%%%%%%%%%%%%%%%%%%%%%%%%%%%%%%%%%%%%%%%%%%%%%%%%%%%%%%%%%%%%
%\vspace{-4 mm}
\begin{center}
\begin{figure}[t] %[!htb] %[h] %
\includegraphics[angle = 0,width=0.450\textwidth]{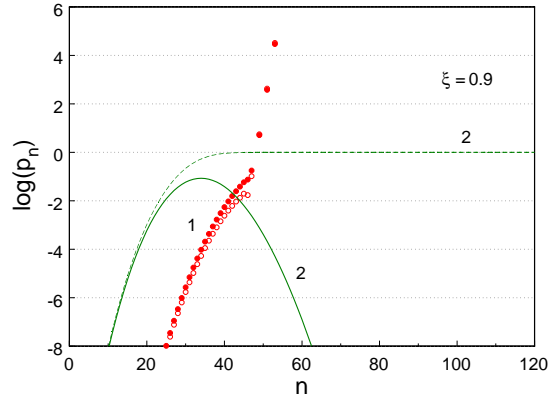}
\caption{(color online)
The probability distribution and the cumulative probability distribution of the number of electron-positron pairs 
in a uniform electric field of supercritical strength $\xi = 0.9$. 
The parameters, except for $\xi$, and the notations are the same as in Fig.~\ref{fig1}. 
The series of $p_n$ converges upon a discrete summation (solid and dashed curves 2). 
In the continuum limit (empty and solid circles 1), 
the series of $p_n$ has an asymptotic character.
}
\label{fig2}
\end{figure}
\end{center}
\vspace{-8mm}
%%%%%%%%%%%%%%%%%%%%%%%%%%%%%%%%%%%%%%%%%%%%%%%%%%%%%%%%%%%%%%%%%%%%%%%%%%%%
%%%%%%%%%%%%%%%%%%%%%%%%%%%%%%%%%%%%%%%%%%%%%%%%%%%%%%%%%%%%%%%%%%%%%%%%%%%%
%%%%%%%%%%%%%%%%%%%%%%%%%%%%%%%%%%%%%%%%%%%%%%%%%%%%%%%%%%%%%%%%%%%%%%%%%%%%

The probability distributions in the continuum limit for $\xi = 0.3$ and $\xi = 0.9$ are shown in Figs.~\ref{fig1} 
and \ref{fig2}. 
The effective number of degrees of freedom is taken to be $\gamma =\gamma _{\Vert } \gamma _{\bot }$, 
with $\gamma_{\Vert } \equiv {|eE|LT}/{\pi } = 10$ and $\gamma _{\bot } \equiv  {|eE|L^{2}}/{(4\pi ^{2})} = 6$. 
In the case $\xi = 0.3$, the distributions are smooth, 
and  with the required accuracy the probabilities sum up to unity. 
In the second case, where $\xi = 0.9$, 
the series has an asymptotic character. 
The apparent divergence arises starting from $ n \sim 50$. 
The order-of-magnitude estimate (\ref{CONS}) suggests that the correlations 
become large for $n \gtrsim n_c \approx 15$.
In the asymptotic region, the figure shows the probability for odd $n$ only. 
The probability for even $n$ turns out negative. 
The violation of positive definiteness occurs when the lower-order alternating terms 
of the recursion (\ref{RECU}) become dominant.
An estimate of the leading alternating term is provided by Eq.~(\ref{ILLU}). 
In the region of $ n$ greater than $50$, a linear semi-log dependence is clearly seen; 
the slope approximately equals $\log(1/r_{\xi}) \sim 1$.
We performed the calculation of $p_n$ in two ways, from the definition (\ref{PRON}) 
and via the recursion (\ref{RECU}).
The results coincide everywhere, including the asymptotic region.

In the asymptotic series, the growing terms are discarded, starting from the smallest one. 
The value of the first term that is omitted is the intrinsic error of calculation. 
In the situation shown in Fig.~\ref{fig2}, the evaluation of the series is not numerically satisfactory: 
Peak values of the probability appear inside the region of divergence, which makes the intrinsic error large; 
nor is the cumulative probability $\sum_{k \leq n}p_k$ saturated 
in the region $n \lesssim 50$. With decreasing $\xi$, the convergence becomes more consistent, 
the asymptotic area moves to the region with high $n$, and, for $\xi < 1/2$, 
the continuum limit finally provides the convergence.

In discrete schemes, the values of $p_n$ are explicitly positive.
In the continuum limit, this property gets lost.
For a supercritical field, this fact has been demonstrated in Fig.~\ref{fig2}.
The probability in the tail of the distribution for $\xi < 1/2$ is also not positive definite.
One can show that the values of $p_n$ in Fig.~\ref{fig1} are negative, 
in the continuum limit, for even $n$ starting from $n=132$.

The probability distribution is well defined for low $n$ in weak fields, 
where $\mathrm{E}[n] =\gamma \xi \ll n_c \sim \gamma$;
the variance is also small in comparison with $n_c$. 
In such cases, the physically interesting values of $n$ lie in the region below $n_c$, 
where the $p_n$ are positive and the continuum limit is efficient.

The results of the discrete summation are also shown in Figs.~\ref{fig1} and \ref{fig2}. 
The solid curves correspond to the probability distribution, whereas the dashed curves show the cumulative 
probability distribution. The discrete distributions are smooth everywhere.
A strong shift of the maximum probability is associated with the replacement of the discrete sum by an integral
and the relatively low value of $\gamma _{\bot } = 6$.
In discrete schemes, the mean value $\mathrm{E}[n]$ and the variance $\mathrm{V} [n]$ can also be expressed in terms of the PGF, 
but because we do not have a closed expression for the PGF of a discrete type, 
this fact does not provide any advantages over the direct calculation. 
Nevertheless, the recursion is numerically efficient. The coefficients $g_l$ 
of a discrete type are as follows:
\[
g_{l}=\frac{(-1)^{l}}{\gamma _{\bot }}\sum\limits_{n=1}^{\infty }\sum\limits_{m=1}^{\infty}
 \left( {\exp (\frac{\pi (n^{2}+m^{2})}{4\gamma _{\bot }})/\xi -1}\right)^{- l - 1}.
\]
In the continuum limit, we arrive at
\[
g_l=(-1)^{l}\int_{0}^{\infty }dx\left( {\exp (x)/\xi -1}\right) ^{- l-1},
\]
which coincides with Eq.~(\ref{GELL}).

The long-wavelength part of the spectrum gives a contribution to $g_{l}$, 
which grows exponentially for large $l$. Accordingly, exponential accuracy is required 
to ensure the cancellation of large terms. In the examples considered, we carry out 
calculations with up to 70 significant digits in order to cover the range of pair numbers $n \leq 120$.
In the subcritical fields, there is no such difficulty.

The proximity to the continuum is controlled by the value of $\gamma _{\bot }$. 
With increasing $\gamma _{\bot }$ and for a constant $\gamma$, curves 2 in Fig.~\ref{fig2} 
approach the initial segments of curves 1, and they remain regular in the region of asymptotic 
divergence of curves 1.

The probability distribution constructed in Secs. II and III corresponds to 
the one-loop QED effective action \cite{HEIS36} the imaginary part of which determines, 
in a uniform electric field, the pair-production rate. 
The two-loop QED effective action and the corresponding correction to pair-production rate
have been calculated by Ritus \cite{RITU75}, Dittrich and Reuter \cite{DITT85} 
and Dittrich and Gies \cite{DITT98}. 
The effect of the two-loop correction on the probability distribution deserves a separate study.

The generalization of our results to an arbitrary spacetime dimension $d + 1$ is straightforward.
The PGF takes the form of Eq.~(\ref{SCHW}), with $\mathrm{Li_2}( (1-z)\xi )$ being
replaced by $\mathrm{Li_s}( (1-z)\xi )$, where $s=(d + 1)/2$. 
For double-degenerate fermions, the coefficient $\gamma$ equals
\[
\gamma = VT\frac{2|eE|^{s}}{(2\pi )^{d}},
\] 
where $V$ is the spatial volume. $\mathrm{Li_s}(t)$ has a branch point at $t=1$, 
which leads to the singularity of the PGF at $z = - r_{\xi}$. 
The cumulative probability converges for $\xi < 1/2$. 
Therefore, the condition of convergence does not depend on the dimension of spacetime.

Recent interest in Klein's paradox was caused by the possibility of 
an experimental study of the production of quasiparticle pairs using
electrostatic barriers in graphene \cite{KATS06}. 
The energy spectrum of quasiparticles in single-layer graphene has a conical
shape, so the quasiparticles are described by the Dirac equation in $2 + 1$ spacetime 
with massless fermions. In single-layer graphene, the branch-point singularity 
of the PGF touches the origin. This gives rise to a strong divergence 
and a failure of the continuum limit approximation.
In bi-layer graphene, quasiparticles have a gapless parabolic spectrum. 
The analytical properties of the PGF are determined by the branch cut that also starts at $z=0$. 

The Schwinger pair production in multilayer graphene is discussed in Refs. \cite{KATS12,ZUBK12,VOLO12}.

The surface of bare strange stars is surrounded by an electron layer 
with a thickness of a few hundred fermi \cite{ALCO85}. The electrons 
cause a strong electric field in the layer. 
If the temperature of the surface is not zero, electron-positron pairs are spontaneously created \cite{USOV98}. 
The effect leads to the radiation of electron-positron plasma, which provides a signature
for the identification of strange stars. 
On the surfaces, the electric field is supercritical, 
so a discrete scheme could be required for quantitative characterization 
of the pair production in strange stars.

The Schwinger mechanism is used in string models for the description of 
decays of orbitally excited hadrons \cite{KOBZ87} and
particle production in high-energy strong interactions \cite{BIRO84}.
The multiplicity of final-state hadrons is determined by the probability distribution of the number of quark-antiquark pairs 
produced in the color electric field of the string.

%%%%%%%%%%%%%%%%%%%%%%%%%%%%%%%%%%%%%%%%%%%%%%%%%%%%%%%%
\section{Conclusion}
%%%%%%%%%%%%%%%%%%%%%%%%%%%%%%%%%%%%%%%%%%%%%%%%%%%%%%%%

In this paper, we found an analytical expression for the PGF in the continuum limit, 
and we used it to calculate the probability distribution, 
as well as the average and variance of the number of electron-positron pairs. 
The probability for large numbers of pairs was calculated using a recursive formula, 
which we have also presented in an explicit analytical form.
If the condition $\xi <1/2$ holds, the branch point of the PGF is outside the unit circle. 
The coefficients $p_n$ of equation (\ref{PRON}), therefore, sum up to unity, 
confirming the meaning of the PGF as a generating function of the probability distribution. 
When $\xi$ becomes greater than 1/2, the branch point of the PGF steps inside the unit circle, 
which causes the sum of the $p_n$ to be divergent. 
This divergence occurs because of an increased contribution of the long-wavelength modes, 
which indicates a failure of the continuum limit approximation.
The probability in the tail of the distribution turns out to be not positive definite, for any field strength. 
The unphysical region %, in which the probability can be negative, 
starts with $n \sim n_c$. 
In the subcritical fields, the cumulative probability is saturated below the $n_c $, 
thus justifying results of the continuum limit approximation. 
In the supercritical fields, the cumulative probability is not necessarily saturated below $ n_c $. 
Whenever this happens, the continuum limit approximation fails.

The PGF, the probability distribution, and the recursion also were calculated in a discrete scheme.
For any field strength, the discrete type PGF is an analytic function inside the unit circle $|z|<1$; 
the coefficients $p_n$ of equation (\ref{PRON}) are positive for any $n$; these coefficients sum up to unity.
In the supercritical fields, discrete summation provides, however, only the basic opportunity to work 
with the convergent series and positive-definite probability. %in the tail of the distributions. 
For quantitative estimates, it is necessary to 
perform calculations with exponential accuracy to guarantee cancellation of large correlation terms.
In the subcritical fields, calculations do not require such high accuracy.

A complete description of the stochastic process is provided by a probability distribution.
Experimental study of the probability distribution of the number of electron-positron pairs  
and/or recursion relations could represent the comprehensive test of the Schwinger mechanism.

\begin{acknowledgments}
The author is grateful to W. Dittrich for reading the manuscript and providing valuable %helpful 
comments and suggestions.
The work was supported in part by the Project No. 3172.2012.2 of Leading Scientific Schools of Russian Federation.
\end{acknowledgments}

\end{document}